\documentclass[12pt,fleqn]{article}

\usepackage{a4wide,epsfig}

\newcommand{\dRRNA}{\frac{d_F^{abcd}d_F^{abcd}}{N_A}}
\newcommand{\dRANA}{\frac{d_F^{abcd}d_A^{abcd}}{N_A}}
\newcommand{\dAANA}{\frac{d_A^{abcd}d_A^{abcd}}{N_A}}

\newcommand{\dRRNF}{\frac{d_F^{abcd}d_F^{abcd}}{N_F}}
\newcommand{\dRANF}{\frac{d_F^{abcd}d_A^{abcd}}{N_F}}

\newcommand{\msbar}{$\overline{\mbox{MS}}$ }

\begin{document}

\title{\vspace{-2cm}
{\small \hfill DESY 04-223 \\
        \hfill SFB/CPP-04-62 \\} \vspace{2cm}
\bf The Four-Loop QCD $\beta$-Function and Anomalous Dimensions}
  
\author{M. Czakon \\ [.5cm]
DESY, Platanenallee 6, D-15738 Zeuthen, Germany \\ [.5cm]
Institute of Physics, University of Silesia, \\
Uniwersytecka 4, PL-40-007 Katowice, Poland}

\maketitle

\begin{abstract}

The four-loop $\beta$-function of quantum chromodynamics is
calculated and agreement is found with the previous result. The 
anomalous dimensions of the quark-gluon vertex, and quark, gluon and
ghost fields are given for a general compact simple Lie group.

\end{abstract}

\section{Introduction}

The renormalization group properties of Quantum Chromodynamics were the
reason of acceptance of this theory as the theory of strong
interactions. The central r\^ole played by the QCD $\beta$-function,
calculated at the one- \cite{1lbeta}, two- \cite{2lbeta}, three-
\cite{3lbeta} and finally at the four-loop \cite{4lbeta} level,
cannot be overestimated in this respect. 

The calculation of the last known, four-loop, term in the
expansion of the $\beta$-function, was performed by only one group
\cite{4lbeta}. The authors evaluated ``...of the order of 50.000
4-loop diagrams''. These two facts lead to the conclusion, stressed
many times (in {\it e.g.} \cite{4lrenormalization}) that it is not
only justified, but also necessary to independently evaluate this
quantity. The present work is intended to fulfill this need.

Apart from the $\beta$-function itself, we present two new
results. The first one is the complete set of \msbar anomalous
dimensions at the four-loop level in the linear gauge and with color
structures of a general compact simple Lie group. These give, in a
compact form, the complete set of four-loop QCD renormalization
constants. We notice that the case of the SU(N) group has been
solved in \cite{4lrenormalization} with the assumption, however, that
the four-loop $\beta$-function is correct. Our second new result is of
a more technical nature, but should simplify future renormalization
group calculations at this level of the perturbative expansion. To
this end, we derived the necessary set of divergent parts of the
four-loop fully massive tadpole master integrals. A purely numeric
result was available from \cite{Laporta:2002pg}, but could not be used
for our purpose.

This paper is organized as follows. The next section presents the
methods used in the calculation, as well as some efficiency
considerations. Subsequently, our results for the anomalous dimensions
are listed. Conclusions and remarks close the main part of the
paper. The expansions of the tadpole master integrals are contained in
the Appendix.

\section{Calculation}

The calculation of renormalization group parameters, {\it i.e.}
anomalous dimensions and beta functions allows for vast
simplifications in comparison to the actual evaluation of Green
functions with kinematic invariants in the physical
region. Dimensional regularization and the \msbar
scheme are particularly well suited for this kind of problems, since they
make it possible to manipulate dimensionful parameters of the
theory. In fact, ever since the introduction of the Infrared
Rearrangement \cite{vladimirov}, it is known how to set most of them
to zero and avoid spurious infrared poles. With the advent of the
$\mbox{R}^*$ operation \cite{rstar}, infrared divergences are even allowed in
individual diagrams and only compensated by counterterms afterward.

At the four-loop level two techniques seem to be most promising. One
is a global $\mbox{R}^*$ operation \cite{4lrenormalization}, where one would
set all of the external momenta to zero and also almost all of the
masses, keeping just one massive line. The spurious infrared
divergences are then compensated by adding a global counterterm. The
advantage of this approach is that the whole problem can be reduced to
the calculation of three-loop massless propagators, for which there
exists a well tested and efficient FORM \cite{form} package, MINCER
\cite{mincer}. The disadvantage is that the construction of the global
counterterm is not trivial. In fact, up to now it has not been possible 
for the gluon propagator.

A second technique consists in setting all of the external momenta to
zero, but keeping a common non-zero mass for all the internal
lines \cite{earlymisiak,4lbeta,Chetyrkin:1997fm}. The advantage of this approach
is that one never encounters any infrared divergences. One minor part
of the price for this convenience is the necessity of a gluon mass
counterterm. The more problematic part is, of course, the calculation
of the divergent parts of four-loop tadpole diagrams that occur. One
way to do this is to generalize the algorithms of
\cite{Chetyrkin:1997fm} to the four-loop level. We, however, decided
to use integration-by-parts identities to reduce all of the integrals
to a set of master integrals depicted in Fig.~\ref{prototypes}. The
divergent parts of the latter were then calculated as described in
Appendix~\ref{expansions}.

\begin{figure}
  \begin{center}
    \epsfig{file=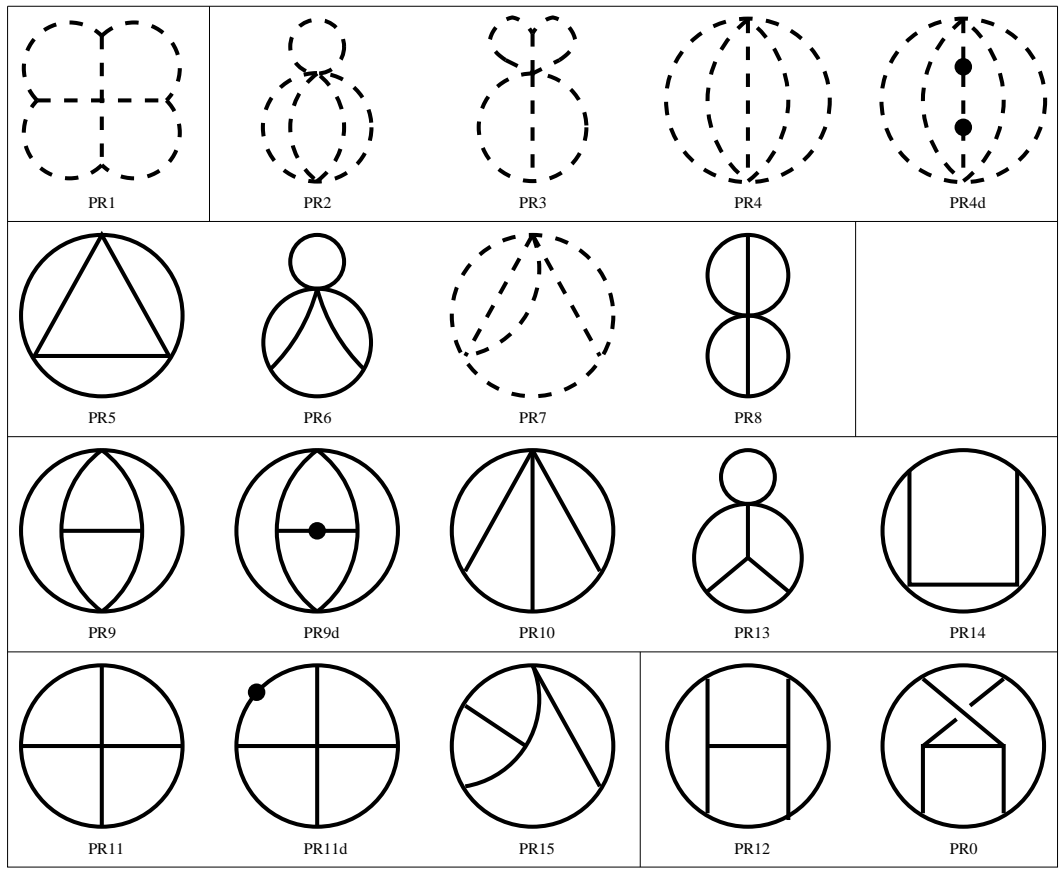,width=14cm}
  \end{center}
  \caption{\label{prototypes} Completely massive four-loop tadpole
  master integrals. Dashed lines mark those integrals which are needed
  to one order higher in the $\epsilon$ expansion, {\it i.e.} up to
  finite parts for this calculation.}
\end{figure}

Instead of developing a dedicated software for the reduction of tadpole
integrals\footnote{Such a software has been developed in \cite{york}.}
we used our own implementation of the Laporta algorithm
\cite{laportaalgo} in the form of the C++ library DiaGen/IdSolver
\cite{idsolver}. We found it also a good opportunity to study the
efficiency of this approach on a large scale problem.

Since the calculation was performed in the linear gauge, the gluon
propagator had the form
\begin{equation}
i D_{\mu\nu}(k) = \frac{i}{k^2}\left(-g_{\mu\nu}+\xi \frac{k_\mu k_\nu}{k^2}\right).
\end{equation}
It is clear that this implies that every power of the gauge
parameter will lead to more powers of the denominators and irreducible
numerators in the integrals. A minimal way to test gauge invariance at
the end is to keep at most a single power of $\xi$, which corresponds
to a first order expansion of the result around the Feynman
gauge. Under this restriction, the calculation involved a little less
than 210.000 independent integrals. The distribution of them among the
four-loop tadpole prototypes in the hardest case of the gluon
propagator is given in Tab.~\ref{distribution}. The remaining cases of
the quark-gluon vertex, and quark and ghost propagators involved about
one-third of these integrals and one power of denominator and
irreducible numerator less for each of the prototypes.

\begin{table}
\begin{center}
  \begin{tabular}{||c|c|c||c|c||}
    \hline
    \hline
    Prototype & number & number & denominator & numerator \\
    & of lines & of integrals & powers & powers \\
    \hline
    PR1 & 4 & 6764 & 7 & 3 \\
    \hline
    PR2 &   & 8575 & 6 & 3 \\
    PR3 & 5 & 19402 & 7 & 4 \\
    PR4 &   & 2659 & 5 & 2 \\
    \hline
    PR5 &   & 8944 & 5 & 3 \\
    PR6 & 6 & 26614 & 7 & 5 \\
    PR7 &   & 15058 & 6 & 4 \\
    PR8 &   & 21528 & 7 & 4 \\
    \hline
    PR9 &   & 15906 & 5 & 4 \\
    PR10& 7 & 28244 & 7 & 5 \\
    PR13&   & 6988 & 6 & 4 \\
    PR14&   & 16157 & 7 & 5 \\
    \hline
    PR11& 8 & 11654 & 5 & 5 \\
    PR15&   & 10973 & 6 & 5 \\
    \hline
    PR12& 9 & 2720 & 5 & 5 \\
    PR0 &   & 1394 & 5 & 5 \\
    \hline
    \hline
  \end{tabular}
\end{center}
\caption{\label{distribution} Distribution between the tadpole
  prototypes of the 203580 different integrals occurring in the
  calculation of the gluon propagator in linear gauge limited to at
  most one power of $\xi$. ``Denominator powers'' denote in fact the
  total number of dots on the lines.}
\end{table}

It turned out that to reduce all of the integrals to masters, it was
sufficient to generate integration-by-parts identities with up to six
additional powers of the denominators and four of the numerators for
all of the integrals up to seven lines, and with five additional powers of
the denominators and four of the numerators for the eight- and
nine-liners. The total number of solved integrals was then about
2.000.000, meaning a 10\% efficiency. About 37\% percent of the
integrals turned out to be finite. These could have been eliminated
from the very beginning by a careful study of divergences. We
convinced ourselves, however, that this would not allow for lower
powers of denominators and numerators in the reduction process,
unless some very involved procedure were used.

\section{Results}

Since the $\beta$-function is, up to normalization, the
anomalous dimension of the coupling constant, it is necessary to
perform the complete renormalization of some vertex. To this end we
chose the quark-gluon interaction, mostly because the calculation of
\cite{4lbeta} involved the ghost and gluon instead.

The renormalization constant of the quark-gluon vertex will
subsequently be denoted by $Z_1$, whereas the renormalization constants
of the quark, gluon and ghost fields by $Z_2$, $Z_3$ and $Z_3^c$
respectively. Even though $Z_3^c$ is not necessary for the present
calculation, we derived it in order to have the complete set of
renormalization constants at the four-loop level.

We will not give the renormalization constants explicitly,
but instead we will limit ourselves to the anomalous dimensions, which
are defined by
\begin{equation}
\gamma = -\mu^2\frac{d \log Z}{d \mu^2},
\end{equation}
where $\mu$ is the 't Hooft unit of mass introduced to keep the
renormalized coupling constant dimensionless. Since we use dimensional
regularization and the \msbar scheme, the renormalization constants
can be expanded as
\begin{equation}
Z = 1+\sum_{i=1}^\infty \frac{z^{(i)}(a_s,\xi)}{\epsilon^i} = 1+\sum_{i=1}^\infty
\sum_{j=1}^{i} a_s^i \frac{z^{(i,j)}(\xi)}{\epsilon^j},
\end{equation}
where $a_s$ is connected to the QCD coupling constant $g$ by
$a_s = \alpha_s/(4 \pi) = g^2/(16\pi^2)$. Using the fact that the
dependence of $Z$ on $\mu$ enters only through $a_s$ and $\xi$,
and that the renormalization constants of the gluon field and of the
gauge parameter are equal, one obtains
\begin{equation}
\label{reconstruction}
(-\epsilon+\beta)a_s \frac{\partial \log Z}{\partial a_s}-\gamma_3 (1-\xi)
  \frac{\partial \log Z}{\partial \xi} = -\gamma,
\end{equation}
where the $\beta$-function is simply equal to the anomalous dimension of
$\alpha_s$, {\it i.e.} $\beta = \gamma_{\alpha_s}$. This implies that 
\begin{equation}
\gamma = a_s \frac{\partial z^{(1)}}{\partial a_s} =
-\sum_{i=0}^\infty a_s^{i+1} \gamma^{(i)},
\end{equation}
where now
\begin{equation}
\gamma^{(i)} = -(i+1) z^{(i+1,1)}.
\end{equation}
Eq.~\ref{reconstruction} can be used in turn to reconstruct the original
renormalization constant from a given anomalous dimension.

Since the quark-gluon vertex, as well as the quark and gluon anomalous
dimensions have already been given up to the three-loop level in the
linear gauge in \cite{3lbeta}, we will not reproduce
them here\footnote{During the course of our calculation we found full
agreement with \cite{3lbeta}. Contrary to the four-loop case, our
three-loop calculation was performed without expansion in the gauge
parameter.} but only give our result for the four-loop anomalous
dimensions with color structures of a general compact simple Lie
group. At this point we stress once more, that the results have been
obtained in a first order expansion around the Feynman gauge

\begin{eqnarray}
\label{g1}
\gamma^{(3)}_1 &=&  C_A C_F T_F^2 n_f^2   \left( \frac{7870}{243} -
\frac{8}{3} \zeta_3 + 24 \zeta_4   \right)
+ C_A C_F^2 T_F n_f   \left( - \frac{797}{18} + 118 \zeta_3 + 36
\zeta_4 + 40 \zeta_5  \right) \nonumber \\
&&+ C_A C_F^3   \left( \frac{5131}{12} + 848 \zeta_3 - 1440 \zeta_5  \right)
+ C_A T_F^3 n_f^3   \left(  - \frac{166}{81} + \frac{32}{9} \zeta_3
\right) \nonumber \\
&&+ C_A^2 C_F T_F n_f   \left( - \frac{104542}{243} + \frac{187}{3}
\zeta_3 - 88 \zeta_4 - 20 \zeta_5 \right) + C_F^2 T_F^2 n_f^2   \left(
\frac{304}{9} - 32 \zeta_3  \right)\nonumber \\ 
&&+ C_A^2 C_F^2   \left( - \frac{23777}{36} - 214 \zeta_3 - 66 \zeta_4 + 790 \zeta_5  \right)
+ C_A^2 T_F^2 n_f^2   \left( \frac{6307}{972} + \frac{94}{3} \zeta_3 -
18 \zeta_4  \right) \nonumber \\
&&+ C_A^3 C_F   \left( \frac{10059589}{15552} - \frac{1489}{24}
\zeta_3 + \frac{173}{4} \zeta_4 - \frac{1865}{12} \zeta_5 \right)+
\frac{280}{81} C_F T_F^3 n_f^3 \nonumber \\
&&+ C_A^3 T_F n_f   \left( - \frac{473903}{7776} - \frac{3311}{24}
\zeta_3 + \frac{387}{8} \zeta_4 + 55 \zeta_5   \right) + C_F^3 T_F n_f
\left( \frac{76}{3} - 64 \zeta_3  \right)\nonumber \\ 
&&+ C_A^4   \left( \frac{350227}{3888} + \frac{2959}{72} \zeta_3 -
\frac{111}{32} \zeta_4 - \frac{5125}{96} \zeta_5 \right) 
+ C_F^4   \left(  - \frac{1027}{8} - 400 \zeta_3 + 640 \zeta_5 \right)
\nonumber \\
&&+ n_f \dRANA   \left(  - 48 \zeta_3 + 60 \zeta_5 \right)
+ \dAANA   \left(  - \frac{21}{8} + \frac{367}{4} \zeta_3 -
\frac{335}{4} \zeta_5 \right)\nonumber \\ 
&& + 128 n_f \dRRNF + \dRANF   \left(  - 66 + 190 \zeta_3 - 170
\zeta_5 \right)
\nonumber \\
+&\xi& \left(
 C_A C_F T_F^2 n_f^2   \left( \frac{1076}{243} - \frac{16}{3} \zeta_3 \right)
+ C_A C_F^2 T_F n_f   \left( \frac{767}{12} - 44 \zeta_3 - 12 \zeta_4
\right ) \right. \nonumber \\ && \left.
+ C_A^2 C_F T_F n_f   \left(  \frac{7423}{243}+ \frac{76}{3} \zeta_3 + \frac{9}{2} \zeta_4  \right)
+ C_A^2 C_F^2   \left(  - 3 + \frac{7}{2} \zeta_3 - 5 \zeta_5  \right)
\right. \nonumber \\ && \left.
+ C_A^2 T_F^2 n_f^2   \left( \frac{1229}{972} - \frac{4}{3} \zeta_3 \right)
+ C_A^3 C_F   \left( - \frac{2127929}{31104} - \frac{1013}{24} \zeta_3
+ \frac{87}{16} \zeta_4 + \frac{665}{24} \zeta_5  \right)
\right. \nonumber \\ && \left.
+ C_A^3 T_F n_f   \left( \frac{35345}{7776} + \frac{37}{3} \zeta_3 + \frac{13}{8} \zeta_4  \right)
+ C_A^4   \left( - \frac{1539403}{62208} - \frac{389}{32} \zeta_3 +
\frac{73}{32} \zeta_4 + \frac{55}{8} \zeta_5  \right)
\right. \nonumber \\ && \left.
+ \dAANA   \left( \frac{9}{16} - \frac{139}{8} \zeta_3 \right)
+ \dRANF   \left(  - 1 - 48 \zeta_3 + 70 \zeta_5 \right)
\right),
\end{eqnarray}

\begin{eqnarray}
\label{g2}
\gamma^{(3)}_2 &=& C_A C_F T_F^2 n_f^2   \left( \frac{6835}{243} +
\frac{112}{3} \zeta_3 \right) + C_A C_F^2 T_F n_f   \left( -
\frac{2407}{36}+44 \zeta_3 + 36 \zeta_4 + 160 \zeta_5  \right)
\nonumber \\ &&+ C_A C_F^3   \left( \frac{5131}{12}+848 \zeta_3 - 1440
\zeta_5   \right) + C_A^2 C_F^2   \left(- \frac{23777}{36}  - 214
\zeta_3 - 66 \zeta_4 + 790 \zeta_5  \right) \nonumber \\ &&+ C_A^2 C_F
T_F n_f   \left( - \frac{1365691}{3888}- \frac{119}{3} \zeta_3 - 25
\zeta_4 - 80 \zeta_5   \right) +\frac{280}{81} C_F T_F^3 n_f^3
\nonumber \\ &&+ C_A^3 C_F   \left(  \frac{10059589}{15552}-
\frac{1489}{24} \zeta_3+\frac{173}{4} \zeta_4    - \frac{1865}{12}
\zeta_5  \right) + C_F^2 T_F^2 n_f^2   \left(   \frac{304}{9}- 32
\zeta_3  \right) \nonumber \\ &&+ C_F^3 T_F n_f   \left(
\frac{76}{3} - 64 \zeta_3 \right) + C_F^4   \left(  - \frac{1027}{8}-
400 \zeta_3 + 640 \zeta_5  \right) + 128 n_f \dRRNF    \nonumber \\
&&+ \dRANF   \left(  - 66 + 190 \zeta_3 - 170 \zeta_5 \right)
\nonumber \\ 
+ &\xi& \left( C_A C_F T_F^2 n_f^2   \left( \frac{1076}{243} -
\frac{16}{3} \zeta_3 \right) + C_A C_F^2 T_F n_f   \left(  \frac{767}{12} - 44
\zeta_3 - 12 \zeta_4  \right) \right. \nonumber \\ &&
\left. + C_A^2 C_F T_F n_f   \left( \frac{48865}{3888} +
\frac{118}{3} \zeta_3 + \frac{15}{2} \zeta_4   \right) + C_A^2 C_F^2   \left(  - 3
+ \frac{7}{2} \zeta_3 - 5 \zeta_5  \right) \right. \nonumber \\ &&
\left.
+ C_A^3 C_F   \left( -
\frac{2127929}{31104} - \frac{1013}{24} \zeta_3  + \frac{87}{16} \zeta_4
+ \frac{665}{24} \zeta_5   \right) \right. \nonumber \\ && \left.
+ \dRANF   \left(  - 1 - 48 \zeta_3 + 70
\zeta_5 \right) \right),
\end{eqnarray}

\begin{eqnarray}
\label{g3}
\gamma^{(3)}_3 &=& C_A C_F T_F^2 n_f^2   \left( - \frac{15082}{243} -
\frac{1168}{9} \zeta_3 + 48 \zeta_4   \right) + C_A C_F^2 T_F n_f   \left(
\frac{10847}{54} + \frac{980}{9} \zeta_3 - 240 \zeta_5
\right)\nonumber \\&& + C_A
T_F^3 n_f^3   \left(  - \frac{1420}{243} + \frac{64}{9} \zeta_3 \right) +
C_A^2 C_F T_F n_f   \left( - \frac{363565}{1944} + \frac{2492}{9}
\zeta_3 - 126 \zeta_4 + 120 \zeta_5   \right) \nonumber \\ && 
+ C_A^2 T_F^2 n_f^2
\left(  - \frac{41273}{486} + \frac{340}{9} \zeta_3 - 36 \zeta_4
\right) + C_F^2 T_F^2 n_f^2   \left(  - \frac{1352}{27} +
\frac{704}{9} \zeta_3 \right) \nonumber \\ &&+
C_A^3 T_F n_f   \left(  \frac{1404961}{3888} - \frac{1285}{4} \zeta_3
+ \frac{387}{4} \zeta_4 + 110 \zeta_5   \right) 
- 46 C_F^3 T_F n_f
\nonumber \\ && + C_A^4   \left( -
\frac{252385}{1944} + \frac{1045}{12} \zeta_3 - \frac{111}{16}
\zeta_4 - \frac{5125}{48} \zeta_5  \right) - \frac{1232}{243} C_F
T_F^3 n_f^3 \nonumber \\ && 
+ n_f \dRANA   \left( -
\frac{512}{9} + \frac{1376}{3} \zeta_3 + 120 \zeta_5 \right) + n_f^2
\dRRNA   \left( \frac{704}{9} - \frac{512}{3} \zeta_3 \right)
\nonumber \\ &&+ \dAANA   \left(
\frac{131}{36} - \frac{307}{6} \zeta_3 - \frac{335}{2} \zeta_5 \right)
\nonumber \\
+&\xi& \left( C_A^2 C_F T_F n_f   \left( \frac{863}{24} - 28 \zeta_3 - 6 \zeta_4  \right)
+ C_A^2 T_F^2 n_f^2   \left( \frac{1229}{486} - \frac{8}{3} \zeta_3 \right) \right.
\nonumber \\ &&\left.
+ C_A^3 T_F n_f   \left( \frac{35345}{3888} + \frac{74}{3} \zeta_3 +
\frac{13}{4} \zeta_4  \right) 
+ C_A^4   \left( - \frac{1539403}{31104} - \frac{389}{16} \zeta_3 +
\frac{73}{16} \zeta_4 + \frac{55}{4} \zeta_5  \right) \right. \nonumber \\
&& \left.
+ \dAANA   \left( \frac{9}{8} - \frac{139}{4} \zeta_3 \right)
\right).
\end{eqnarray}

The first three terms of the expansion of the anomalous dimension of
the ghost field in the linear gauge can be found in
\cite{4lrenormalization}. Even though the results there are restricted
to the SU(N) group, the values for a general compact simple Lie group
can be derived on the basis of the fact that only quadratic Casimir
operators occur.

As before our four-loop result is obtained in a first order expansion
around the Feynman gauge

\begin{eqnarray}
\label{g3c}
\gamma^{c\; (3)}_3 &=& C_A C_F T_F^2 n_f^2   \left( - \frac{115}{27} + 40
\zeta_3 - 24 \zeta_4  \right) + C_A C_F^2 T_F n_f   \left( -
\frac{271}{12} - 74 \zeta_3 + 120 \zeta_5  \right) \nonumber \\
&&+ C_A T_F^3 n_f^3
\left( \frac{166}{81} - \frac{32}{9} \zeta_3 \right) 
+C_A^2 T_F^2 n_f^2   \left( - \frac{8315}{972} - \frac{86}{3} \zeta_3
+ 18 \zeta_4   \right) \nonumber \\ 
&&+ C_A^2 C_F T_F n_f
\left( \frac{22517}{432} - 86 \zeta_3 + 69 \zeta_4 - 60 \zeta_5
\right) \nonumber \\
&&+ C_A^3 T_F n_f   \left( \frac{449239}{7776} +
\frac{2983}{24} \zeta_3 - \frac{423}{8} \zeta_4 - 55 \zeta_5  \right)
\nonumber \\
&&+ C_A^4   \left( - \frac{256337}{3888}- \frac{2485}{72} \zeta_3 +
\frac{123}{32} \zeta_4 + \frac{4505}{96} \zeta_5  \right) \nonumber \\
&&+ n_f \dRANA
\left( 48 \zeta_3 - 60 \zeta_5 \right) + \dAANA   \left( \frac{21}{8} -
\frac{299}{4} \zeta_3 + \frac{265}{4} \zeta_5  \right) 
\nonumber \\
+ &\xi& \left( C_A^2 C_F T_F n_f   \left( \frac{425}{48} - 2 \zeta_3
- 3 \zeta_4  \right) + C_A^2 T_F^2 n_f^2   \left( \frac{779}{972} -
\frac{4}{3} \zeta_3 \right) \right. \nonumber \\ && \left.
+ C_A^3 T_F n_f   \left(  - \frac{2527}{7776} +
\frac{7}{4} \zeta_3 + \frac{23}{8} \zeta_4 \right) + C_A^4   \left(  -
\frac{256273}{62208} + \frac{199}{48} \zeta_3  - \frac{47}{16}
\zeta_4 + \frac{25}{48} \zeta_5 \right) \right. \nonumber \\ && \left.
+ \dAANA   \left(  - \frac{9}{16} -
\frac{27}{8} \zeta_3 + \frac{85}{4} \zeta_5 \right) \right).
\end{eqnarray}

The notation is similar to the one used in \cite{4lbeta}. Besides Riemann
$\zeta$ functions, the result contains the quadratic Casimir operators
of the fundamental and adjoint representations, $C_F$ and $C_A$, as
well as the normalization of the trace of the fundamental
representation $T_F\; \delta^{ab}=\mbox{Tr} \left(T^a T^b\right)$, where $T^a$ are the
representation generators, and the number of fermion families
$n_f$. The higher order invariants are constructed from the symmetric
tensors
\begin{eqnarray}
d_F^{abcd} &=& \frac{1}{6} \mbox{Tr}\left( T^a T^b T^c T^d
+T^a T^b T^d T^c
+T^a T^c T^b T^d \right. \nonumber \\ && \;\;\;\; \left.
+T^a T^c T^d T^b
+T^a T^d T^b T^c
+T^a T^d T^c T^b \right),
\end{eqnarray}
(and similarly for the adjoint representation) and of the dimensions
of the fundamental and adjoint representations, $N_F$ and $N_A$
respectively. Using the specific values for the SU(N) group
\begin{equation}
T_F = \frac{1}{2}, \;\;\;\; C_F = \frac{N^2-1}{2N}, \;\;\;\; C_A =N,
\;\;\;\; \dRRNA = \frac{N^4-6N^2+18}{96N^2},
\end{equation}
\begin{equation}
\dRANA = \frac{N(N^2+6)}{48}, \;\;\;\; \dAANA =
\frac{N^2(N^2+36)}{24}, \;\;\;\; N_A = N^2-1,
\end{equation}
we checked that Eqs.~\ref{g1}-\ref{g3c} are in perfect agreement with
\cite{4lrenormalization}.

We can now combine the anomalous dimensions to reach the goal of our
calculation, {\it i.e.} the four-loop $\beta$-function. Since,
$Z_{\alpha_s} = Z^2_g = Z_1^2 Z_2^{-2} Z_3^{-1}$, we have $\beta =
2\gamma_1-2\gamma_2-\gamma_3$, and
\begin{eqnarray}
\beta_3 &=& C_A C_F T_F^2 n_f^2   \left( \frac{17152}{243} + \frac{448}{9} \zeta_3  \right)
+ C_A C_F^2 T_F n_f   \left( - \frac{4204}{27} + \frac{352}{9} \zeta_3  \right)
+ \frac{424}{243} C_A T_F^3 n_f^3 \nonumber \\
&&+ C_A^2 C_F T_F n_f   \left( \frac{7073}{243} - \frac{656}{9} \zeta_3 \right)
+ C_A^2 T_F^2 n_f^2   \left( \frac{7930}{81} + \frac{224}{9} \zeta_3
\right)
+  \frac{1232}{243} C_F T_F^3 n_f^3 \nonumber \\
&&+ C_A^3 T_F n_f   \left( - \frac{39143}{81} + \frac{136}{3} \zeta_3  \right)
+ C_A^4   \left(  \frac{150653}{486} - \frac{44}{9} \zeta_3  \right)
+ C_F^2 T_F^2 n_f^2   \left( \frac{1352}{27} - \frac{704}{9} \zeta_3
\right) \nonumber \\
&&+ 46 C_F^3 T_F n_f
+ n_f \dRANA   \left( \frac{512}{9} - \frac{1664}{3} \zeta_3 \right)
+ n_f^2 \dRRNA   \left(  - \frac{704}{9} + \frac{512}{3} \zeta_3
\right) \nonumber \\
&&+ \dAANA   \left(  - \frac{80}{9} + \frac{704}{3} \zeta_3 \right).
\end{eqnarray}
This result, manifestly gauge invariant, confirms \cite{4lbeta}. For
completeness, we reproduce the lower order values, which we have also calculated
\begin{eqnarray}
\beta_2 &=& \frac{2857}{54} C_A^3  - \frac{1415}{27} C_A^2 T_F n_f
+ \frac{158}{27} C_A T_F^2 n_f^2+ \frac{44}{9} C_F T_F^2 n_f^2
\nonumber \\
&&- \frac{205}{9} C_F C_A T_F n_f+ 2 C_F^2 T_F n_f,
\nonumber \\
\beta_1 &=& \frac{34}{3} C_A^2 - \frac{20}{3} C_A T_F n_f- 4 C_F
T_F n_f,
\nonumber \\
\beta_0 &=& \frac{11}{3} C_A - \frac{4}{3} T_F n_f.
\end{eqnarray}

\section{Conclusions}

We have completed the four-loop \msbar renormalization program of an
unbroken gauge theory with fermions and a general compact simple Lie
group in the linear gauge. The fact that we performed  an expansion in
the gauge parameter still allows for gauge invariance tests in
practical calculations, although special gauges such as the Landau
gauge cannot be chosen. If such need would occur, one may use the
SU(N) results from \cite{4lrenormalization}. The correctness of the
latter relies, however, on the four-loop $\beta$-function. Therefore,
the present work was not only indispensable to confirm the value of
the four-loop $\beta$-function itself, but also the correctness of
\cite{4lrenormalization}. This goal has been reached.

It should be stressed that with the accumulated solved integrals, one
could easily obtain one more term in the $\xi$ expansion of the
anomalous dimensions. This is due
to the fact that the complexity of the integrals occurring in the gluon
propagator with at most a single power of $\xi$ is comparable to the complexity
of the integrals for the other functions with at most two powers of
$\xi$. Having solved the latter, the gauge invariant $\beta$-function
enables then to recover the third term in the $\xi$ expansion of the gluon
propagator as well. We did not perform such a calculation, since from the
purely pragmatical point of view, only the complete $\xi$ dependence
would be an improvement.

A second comment concerns the independence of our calculation. The
reader should notice that besides basic software as FORM \cite{form},
Fermat \cite{fermat} {\it etc.} and our own programs \cite{idsolver},
the only external package that we used was the Color package \cite{color}
of the FORM distribution. On a diagram per diagram basis, we made sure
that the color factors produced by this package agree with the SU(N)
values obtained with the algorithm of \cite{cvitanovich}. Our
confidence was strengthened by the fact that we agreed with
\cite{4lrenormalization} on the final results.

\section {Acknowledgements}

The author would like to thank K. Chetyrkin, M. Misiak and
Y. Schr\"oder for stimulating discussions, and the Institute of
Theoretical Physics and Astrophysics of the University of W\"urzburg
for warm hospitality during the period when part of this work was
completed.  This work was supported in part by TMR, European Community
Human Potential Programme under contract HPRN-CT-2002-00311
(EURIDICE), by Deutsche Forschungsgemeinschaft  under  contract SFB/TR
9--03, and by the Polish State Committee for Scientific Research (KBN)
under contract No. 2P03B01025.

\appendix

\section{Master integral expansions}

\label{expansions}

Here we give the $\epsilon$ expansions of the master integrals depicted in
Fig.~\ref{prototypes} up to the necessary order. The notation is
similar to that used in MATAD \cite{Steinhauser:2000ry}, {\it i.e.} we
use the integration measure
\begin{equation}
\left( e^{\epsilon \gamma_E} \right)^4\int 
\prod_{i=1}^{4}\frac{d^d k_i}{i\pi^{d/2}},
\end{equation}
with $d=4-2\epsilon$ and the constants which occur in the divergent
parts of the integrals are
\begin{eqnarray}
\mbox{S2} &=& \frac{4}{9\sqrt{3}}\mbox{Cl}_2\left(\frac{\pi}{3}\right),
\\
\mbox{T1ep} &=& -\frac{45}{2}-\frac{\pi\sqrt{3}\log^2 3}{8}-\frac{35\pi^3\sqrt{3}}{216}
        -\frac{9}{2}\zeta_2+\zeta_3 \nonumber\\
	&+&6\sqrt{3}\mbox{Cl}_2\left(\frac{\pi}{3}\right)
        -6\sqrt{3}\mbox{Im}\left(\mbox{Li}_3
	\left(\frac{e^{-i\frac{\pi}{6}}}{\sqrt{3}}\right)\right), 
	\nonumber \\
\mbox{D6} &=& 6\zeta_3-17\zeta_4-4\zeta_2\log^2 2+\frac{2}{3}\log^4 2
      +16\mbox{Li}_4\left(\frac{1}{2}\right)-4\left(\mbox{Cl}_2\left(\frac{\pi}{3}\right)\right)^2,
      \nonumber
\end{eqnarray}
where $\mbox{Cl}_2(x)=\mbox{Im}(\mbox{Li}_2(e^{i x}))$ is the Clausen
function. Notice, however, that we use the Minkowski space metric and
``Minkowski type'' propagators, {\it i.e.} $1/(k^2-m^2)$.

\begin{eqnarray*}
\mbox{\bf PR0} &=& {\cal O}(\epsilon^0),
\end{eqnarray*}

\begin{eqnarray*}
\mbox{\bf PR12} &=& {\cal O}(\epsilon^0), 
\end{eqnarray*}

\begin{eqnarray*}
\mbox{\bf PR15} &=& \frac{1}{\epsilon^2}\;\frac{3}{2}\; \zeta_3
+\frac{1}{\epsilon}\;\left(\mbox{D6}+\frac{3}{2}\;\zeta_3-\frac{3}{4}\;\zeta_4\right)
+{\cal O}(\epsilon^0),
\end{eqnarray*}

\begin{eqnarray*}
\mbox{\bf PR11} &=& \frac{5}{\epsilon}\;\zeta_5+{\cal O}(\epsilon^0),
\end{eqnarray*}

\begin{eqnarray*}
\mbox{\bf PR11d} &=& {\cal O}(\epsilon^0),
\end{eqnarray*}

\begin{eqnarray*}
\mbox{\bf PR14} &=& \frac{1}{\epsilon^4}\; \frac{3}{4}
+\frac{1}{\epsilon^3}\; \frac{25}{4}
+\frac{1}{\epsilon^2}\;
\left(\frac{137}{4}-\frac{81}{2}\;\mbox{S2}+\frac{3}{2}\;\zeta_2
\right) \nonumber \\
&&+\frac{1}{\epsilon}\left(\frac{363}{4}-162\;
\mbox{S2}-3\;\mbox{T1ep}-\zeta_2-\frac{27}{2}\;\zeta_3\right)
+{\cal O}(\epsilon^0),
\end{eqnarray*}

\begin{eqnarray*}
\mbox{\bf PR13} &=&
\frac{2}{\epsilon^2}\;\zeta_3+\frac{1}{\epsilon}(\mbox{D6}+2\zeta_3)+{\cal O}(\epsilon^0),
\end{eqnarray*}

\begin{eqnarray*}
\mbox{\bf PR10} &=& \frac{1}{\epsilon^4}\;\frac{1}{2}
+\frac{1}{\epsilon^3}\; \frac{53}{12}
+\frac{1}{\epsilon^2}\; \left(\frac{51}{2}-27\;
\mbox{S2}+\zeta_2\right) \nonumber \\
&&+\frac{1}{\epsilon}\;\left(\frac{937}{12}-
\frac{243}{2}\;\mbox{S2}-2\;\mbox{T1ep}-\frac{1}{6}\;\zeta_2-\frac{32}{3}\;\zeta_3\right)
+{\cal O}(\epsilon^0),
\end{eqnarray*}

\begin{eqnarray*}
\mbox{\bf PR9} &=& \frac{1}{\epsilon^4}\;\frac{1}{4}
+\frac{1}{\epsilon^3}\;\frac{7}{3}
+\frac{1}{\epsilon^2}\;\left(\frac{169}{12}-\frac{27}{2}\;\mbox{S2}+\frac{1}{2}\;\zeta_2
+\zeta_3\right) \nonumber \\
&&+\frac{1}{\epsilon}\left(\frac{143}{3}-\frac{135}{2}\;\mbox{S2}
-\mbox{T1ep}+\frac{1}{6}\;\zeta_2-\frac{4}{3}\;\zeta_3+\frac{3}{2}\;\zeta_4\right)
+{\cal O}(\epsilon^0),
\end{eqnarray*}

\begin{eqnarray*}
\mbox{\bf PR9d} &=& \frac{1}{\epsilon^4}\;\frac{1}{12}
+\frac{1}{\epsilon^3}\;\frac{1}{3}
+\frac{1}{\epsilon^2}\;\left(\frac{7}{12}-\frac{9}{2}\;\mbox{S2}+\frac{1}{6}\;\zeta_2\right)
\nonumber \\ &+&\frac{1}{\epsilon}\;\left(-\frac{26}{3}+\frac{27}{2}\;\mbox{S2}
-\frac{1}{3}\;\mbox{T1ep}-\frac{5}{6}\;\zeta_2+\frac{29}{9}\;\zeta_3\right)
+{\cal O}(\epsilon^0),
\end{eqnarray*}

\begin{eqnarray*}
\mbox{\bf PR8} &=& \frac{1}{\epsilon^4}\;\frac{9}{4}
+\frac{1}{\epsilon^3}\;\frac{27}{2}
+\frac{1}{\epsilon^2}\;\left(\frac{207}{4}-\frac{81}{2}\;\mbox{S2}
+\frac{9}{2}\;\zeta_2\right) \nonumber \\
&&+\frac{1}{\epsilon}\left(\frac{189}{2}-\frac{243}{2}\;\mbox{S2}
-3\;\mbox{T1ep}+\frac{27}{2}\;\zeta_2\right)
+{\cal O}(\epsilon^0),
\end{eqnarray*}

\begin{eqnarray*}
\mbox{\bf PR7} &=& \frac{1}{\epsilon^4}\;\frac{13}{8}
+\frac{1}{\epsilon^3}\;\frac{491}{48}
+\frac{1}{\epsilon^2}\;\left(\frac{3719}{96}-\frac{81}{4}\;\mbox{S2}+\frac{13}{4}\;\zeta_2\right)  
\nonumber \\ 
&&+\frac{1}{\epsilon}\;\left(\frac{13741}{192}
-\frac{459}{8}\;\mbox{S2}-\frac{3}{2}\;\mbox{T1ep}+\frac{329}{24}\;\zeta_2
-\frac{2}{3}\;\zeta_3\right) \nonumber \\
&&+\frac{381313}{10368}-\frac{1593}{16}\;\mbox{S2}-\frac{17}{4}\;\mbox{T1ep}
-\frac{3}{2}\;\mbox{T1ep2}-\frac{5}{4}\;\mbox{PR4dfin} \nonumber \\
&&-\frac{1}{4}\;\mbox{PR4fin} 
+\frac{6805}{144}\;\zeta_2-\frac{81}{4}\;\mbox{S2}\;\zeta_2
+\frac{61}{2}\;\zeta_3+\frac{153}{16}\;\zeta_4
+{\cal O}(\epsilon),
\end{eqnarray*}

\begin{eqnarray*}
\mbox{\bf PR6} &=& \frac{1}{\epsilon^4}
+\frac{1}{\epsilon^3}\;\frac{20}{3}
+\frac{1}{\epsilon^2}\left(29-27\;\mbox{S2}+2\;\zeta_2\right) \nonumber \\
&&+\frac{1}{\epsilon}\left(\frac{181}{3}-81\;\mbox{S2}-2\;\mbox{T1ep} 
+\frac{13}{3}\;\zeta_2-\frac{16}{3}\;\zeta_3\right)
+{\cal O}(\epsilon^0),
\end{eqnarray*}

\begin{eqnarray*}
\mbox{\bf PR5} &=& \frac{1}{\epsilon^4}\;\frac{3}{2}
+\frac{1}{\epsilon^3}\;\frac{19}{2}
+\frac{1}{\epsilon^2}\;\left(\frac{67}{2}+3\;\zeta_2\right)
+\frac{1}{\epsilon}\left(\frac{127}{2}+19\;\zeta_2-5\;\zeta_3\right)
+{\cal O}(\epsilon^0),
\end{eqnarray*}

\begin{eqnarray*}
\mbox{\bf PR4} &=& \frac{1}{\epsilon^4}\;\frac{5}{2}
+\frac{1}{\epsilon^3}\;\frac{35}{3}
+\frac{1}{\epsilon^2}\left(\frac{4565}{144}+5\;\zeta_2\right)
\nonumber \\
&&+ \frac{1}{\epsilon}\left(\frac{58345}{864}+\frac{70}{3}\;\zeta_2-\frac{10}{3}\;\zeta_3\right)
+\mbox{PR4fin}+{\cal O}(\epsilon),
\end{eqnarray*}

\begin{eqnarray*}
\mbox{\bf PR4d} &=& -\frac{1}{\epsilon^3}\;\frac{7}{6}
-\frac{1}{\epsilon^2}\;\frac{215}{48}
+\frac{1}{\epsilon}\;\left(-\frac{965}{96}-\frac{7}{3}\;\zeta_2\right)
+\mbox{PR4dfin}+{\cal O}(\epsilon),
\end{eqnarray*}

\begin{eqnarray*}
\mbox{\bf PR3} &=& \frac{1}{\epsilon^4}\;\frac{3}{2}
+\frac{1}{\epsilon^3}\;\frac{15}{2}
+\frac{1}{\epsilon^2}\;\left(24-\frac{27}{2}\;\mbox{S2}+3\;\zeta_2\right)
\nonumber \\
&&+\frac{1}{\epsilon}\;\left(\frac{81}{2}-27\;\mbox{S2}-\mbox{T1ep}+\frac{21}{2}\;\zeta_2
-\zeta_3\right)
+57-\frac{81}{2}\;\mbox{S2}-2\;\mbox{T1ep} \nonumber \\
&&-\mbox{T1ep2}+\frac{57}{2}\;\zeta_2
-\frac{27}{2}\;\mbox{S2}\;\zeta_2-5\;\zeta_3+\frac{51}{8}\;\zeta_4
+{\cal O}(\epsilon),
\end{eqnarray*}

\begin{eqnarray*}
\mbox{\bf PR2} &=& \frac{2}{\epsilon^4}
+\frac{1}{\epsilon^3}\;\frac{29}{3}
+\frac{1}{\epsilon^2}\;\left(\frac{163}{6}+4\;\zeta_2\right)
+\frac{1}{\epsilon}\;\left(\frac{601}{12}+\frac{58}{3}\;\zeta_2-\frac{8}{3}\;\zeta_3\right)
\nonumber \\
&&+\frac{635}{24}+\frac{163}{3}\;\zeta_2+\frac{220}{9}\;\zeta_3+12\;\zeta_4
+{\cal O}(\epsilon),
\end{eqnarray*}

\begin{eqnarray*}
\mbox{\bf PR1} &=& \frac{1}{\epsilon^4}
+\frac{4}{\epsilon^3}
+\frac{1}{\epsilon^2}\;\left(10+2\;\zeta_2\right)
+\frac{1}{\epsilon}\;\left(20+8\;\zeta_2-\frac{4}{3}\;\zeta_3\right)\nonumber \\
&&+35+20\;\zeta_2-\frac{16}{3}\;\zeta_3+6\;\zeta_4
+{\cal O}(\epsilon).
\end{eqnarray*}

The expansions of the irreducible four-loop integrals have been
obtained either from the finiteness of the same integral with higher
powers of denominators or with the help of the method described in
\cite{Chetyrkin:1997fm}. After taking into account the different
normalization and translating master integrals with dots to integrals
with suitable irreducible numerators, the above results for the
divergent parts are in agreement with the numerical values obtained in
\cite{Laporta:2002pg}.

The integrals PR1-PR4, PR4d and PR7 are needed up to constant parts. It is,
however, sufficient to express the finite part of PR7 through the finite
parts of PR4 and PR4d and keep them as symbols, since they
always cancel from the divergent part of any four-loop vacuum integral.

The reader should notice that we also needed $\epsilon$ expansions of
two- and three-loop tadpoles. These have been obtained in
\cite{2ltadpoles,3ltadpoles}. The values can be read off from our
results on the reducible four-loop integrals above.


\begin{thebibliography}{00}

\bibitem{1lbeta}
D.~J.~Gross and F.~Wilczek,
Phys.\ Rev.\ Lett.\  {\bf 30} (1973) 1343;

H.~D.~Politzer,
Phys.\ Rev.\ Lett.\  {\bf 30} (1973) 1346;

G. `t Hooft, report at the Marseille Conference on Yang-Mills Fields, 1972.

\bibitem{2lbeta}

W.~E.~Caswell,
Phys.\ Rev.\ Lett.\  {\bf 33} (1974) 244;

D.~R.~T.~Jones,
Nucl.\ Phys.\ B {\bf 75} (1974) 531;

E.~Egorian and O.~V.~Tarasov,
Theor.\ Math.\ Phys.\  {\bf 41} (1979) 863
[Teor.\ Mat.\ Fiz.\  {\bf 41} (1979) 26].

\bibitem{3lbeta}

O.~V.~Tarasov, A.~A.~Vladimirov and A.~Y.~Zharkov,
Phys.\ Lett.\ B {\bf 93} (1980) 429;

S.~A.~Larin and J.~A.~M.~Vermaseren,
Phys.\ Lett.\ B {\bf 303} (1993) 334.

\bibitem{4lbeta}

T.~van Ritbergen, J.~A.~M.~Vermaseren and S.~A.~Larin,
Phys.\ Lett.\ B {\bf 400} (1997) 379.

\bibitem{4lrenormalization}

K.~G.~Chetyrkin,
arXiv:hep-ph/0405193.

\bibitem{Laporta:2002pg}
S.~Laporta,
Phys.\ Lett.\ B {\bf 549}, 115 (2002).

\bibitem{vladimirov}

A.~A.~Vladimirov,
Theor.\ Math.\ Phys.\  {\bf 43} (1980) 417
[Teor.\ Mat.\ Fiz.\  {\bf 43} (1980) 210].

\bibitem{rstar}

K.~G.~Chetyrkin and V.~A.~Smirnov,
Phys.\ Lett.\ B {\bf 144} (1984) 419;

K.~G.~Chetyrkin,
Phys.\ Lett.\ B {\bf 390} (1997) 309;

K.~G.~Chetyrkin,
Phys.\ Lett.\ B {\bf 391} (1997) 402.

\bibitem{form}

J.~A.~M.~Vermaseren,
arXiv:math-ph/0010025.

\bibitem{mincer}

S.~A.~Larin, F.~V.~Tkachov and J.~A.~M.~Vermaseren,
NIKHEF-H-91-18.

\bibitem{earlymisiak}

M.~Misiak and M.~Munz,
Phys.\ Lett.\ B {\bf 344} (1995) 308.

\bibitem{Chetyrkin:1997fm}

K.~G.~Chetyrkin, M.~Misiak and M.~Munz,
Nucl.\ Phys.\ B {\bf 518} (1998) 473.

\bibitem{york}

Y.~Schr\"oder,
Nucl.\ Phys.\ Proc.\ Suppl.\  {\bf 116} (2003) 402.

\bibitem{laportaalgo}

S.~Laporta,
Int.\ J.\ Mod.\ Phys.\ A {\bf 15} (2000) 5087.

\bibitem{idsolver}

M.~Czakon, DiaGen/IdSolver, {\it unpublished}.

\bibitem{fermat}

R.~H.~Lewis, Fermat, http://www.bway.net/\verb!~!lewis/.

\bibitem{color}

T.~van Ritbergen, A.~N.~Schellekens and J.~A.~M.~Vermaseren,
Int.\ J.\ Mod.\ Phys.\ A {\bf 14} (1999) 41.

\bibitem{cvitanovich}

P.~Cvitanovic,
Phys.\ Rev.\ D {\bf 14} (1976) 1536.

\bibitem{Steinhauser:2000ry}
M.~Steinhauser,
Comput.\ Phys.\ Commun.\  {\bf 134} (2001) 335.

\bibitem{2ltadpoles}

A.~I.~Davydychev and J.~B.~Tausk,
Nucl.\ Phys.\ B {\bf 397} (1993) 123;

A.~I.~Davydychev and J.~B.~Tausk,
Phys.\ Rev.\ D {\bf 53} (1996) 7381;

A.~I.~Davydychev,
Phys.\ Rev.\ D {\bf 61} (2000) 087701.

\bibitem{3ltadpoles}

D.~J.~Broadhurst,
Z.\ Phys.\ C {\bf 54} (1992) 599;

D.~J.~Broadhurst,
Eur.\ Phys.\ J.\ C {\bf 8} (1999) 311;

J.~Fleischer and M.~Y.~Kalmykov,
Phys.\ Lett.\ B {\bf 470} (1999) 168;

K.~G.~Chetyrkin and M.~Steinhauser,
Nucl.\ Phys.\ B {\bf 573} (2000) 617.

\end{thebibliography}
\end{document}